\documentclass[10pt,aps,twocolumn,eqsecnum,nofootinbib,amsmath,amssymb,floatfix]{revtex4}
\usepackage{graphicx}
\usepackage{dcolumn}
\usepackage{bm}
\usepackage{epsfig,psfrag,amsmath,amssymb,float}
\input{epsf}

\begin{document}


\def\leftrh{James B.~Hartle}
\def\rightrh{Quasiclassical Realms In A Quantum Universe}

\def\frac#1#2{{\textstyle #1 \over \textstyle #2}}
\def\half{{\textstyle {1 \over 2}}}
\def\viz{{\it viz.~}}
\def\eg{{\it e.g.~}}
\def\ie{{\it i.e.,\ }}
\def\gtwid{\mathrel{\raise.3ex\hbox{$>$\kern-.75em\lower1ex\hbox{$\sim$}}}}
\def\ltwid{\mathrel{\raise.3ex\hbox{$<$\kern-.75em\lower1ex\hbox{$\sim$}}}}
\def\ed{{\it ed.~by\ }}

\title{Quasiclassical Realms In A Quantum Universe\footnote{Talk
given at the Lanczos Centenary Meeting, North Carolina State University,
December 15, 1993}}

\author{James B.~Hartle}
\email{hartle@physics.ucsb.edu}
\affiliation{Department of Physics,
University of California\\
 Santa Barbara, CA 93106-9530 USA}

\begin{abstract}

In this universe, governed fundamentally by quantum mechanical laws,
characterized by indeterminism and distributed probabilities, classical
deterministic laws are applicable over a wide range of time, place, and
scale.  We review the origin of these deterministic laws in the context
of the quantum mechanics of closed systems, most generally, the universe
as a whole. In this formulation of quantum mechanics, probabilities are predicted for the individual members of
sets of alternative histories of the universe that decohere, \ie for
which there is negligible interference between pairs of histories in the set as measured
by a decoherence functional.  An expansion of the decoherence functional
in the separation between histories allows the form of the
phenomenological, deterministic equations of motion to be derived for 
suitable coarse grainings of a
class of non-relativistic systems, including ones with general non-linear
interactions.  More coarse graining is
needed to achieve classical predictability than naive arguments based on
the uncertainty principle would suggest.  Coarse graining is needed to
effect decoherence, and coarse graining beyond that to achieve the
inertia necessary to resist the noise that mechanisms of decoherence
produce.  Sets of histories governed largely by deterministic laws
constitute the quasiclassical realm of everyday experience which is an
emergent feature of the closed system's initial condition and
Hamiltonian.  We analyse the question of the sensitivity of the
existence of a quasiclassical realm to the particular form of the
initial condition.  We find that almost any initial condition will
exhibit a quasiclassical realm of some sort, but only a small fraction
of the total number of possible initial states could reproduce the everyday
quasiclassical realm of our universe.
\end{abstract}

\maketitle

\setcounter{footnote}{0}
\section{Introduction}
\label{sec:I}

In cosmology we confront a problem which is fundamentally
different from that encountered elsewhere in physics.  This is the
problem of providing a theory of the initial condition of the universe.
The familiar laws of physics describe evolution in time.  The
evolution of a plasma is described by the classical laws of
electrodynamics and mechanics and the evolution of an atomic state by
Schr\"odinger's equation.
These dynamical laws require boundary
conditions and the laws which govern the evolution of the universe ---
the classical Einstein equation, for instance  --- are no exception.  There are
no particular laws governing these boundary conditions; they summarize
our observations of the universe outside the subsystem whose evolution
we are studying.  If we don't see any radiation coming into a room,
then we
solve Maxwell's equations inside with no-incoming-radiation boundary
conditions.  If we prepare an atom in a certain way, then we solve
Schr\"odinger's equation with the corresponding initial condition.

In cosmology, however, by definition, there is no rest of the universe
to pass the specification of the boundary conditions off to.  The
boundary conditions must be part of the laws of physics themselves.
Constructing a theory of the initial condition of the universe,
effectively its initial quantum state, and examining its observational
consequences is the province of that area of astrophysics that has come
to be called quantum cosmology.\footnote{For a recent review 
see \cite{Hal91}}
  This talk will consider one
manifest feature of the quantum universe and its connection to the  
theory of the initial condition.  This is the applicability of the 
deterministic laws of
classical physics to a wide range of phenomena in the universe ranging
from the cosmological expansion itself to the turbulent and viscous flow
of water through a pipe. This quasiclassical realm\footnote{Earlier 
work, e.g. \cite{GH90a} called this the `quasiclassical domain', but 
this risks confusion the usage in condensed matter physics.} 
 is one of the most immediate facts of our
experience. Yet what we know of the basic laws of physics suggests that
we live in a quantum mechanical universe, characterized by indeterminacy
and distributed probabilities, where classical laws can be but
approximations to the unitary evolution of the Schr\"odinger equation
and the reduction of the wave packet.  What is the origin of this wide
range of time, place, and scale on which classical determinism 
applies? How can we derive the form of the phenomenological classical
laws, say the Navier-Stokes equations, from a distantly related
fundamental quantum mechanical
theory which might, after all, be heterotic, superstring
theory? What features of these laws can be traced to their quantum
mechanical origins?  It is such old questions that will be examined
anew in this lecture from the perspective of quantum cosmology, reporting
largely on joint work with Murray Gell-Mann \cite{GH93a}.

Standard derivations of classical behavior from the laws of quantum
mechanics are available in many quantum mechanics texts.  One popular
approach is based on Ehrenfest's theorem relating the acceleration of
the expected value of position to the expected value of the force:
\begin{equation}
m\ \frac{d^2\langle x\rangle}{dt^2} = - \left\langle\frac{\partial
V}{\partial x}\right\rangle\ ,
\label{oneone}
\end{equation}
(written here for one-dimensional motion).  Ehrenfest's
theorem is true in general, but for certain states, typically narrow
wave packets, we may approximately replace the expected value of the
force with the force evaluated at the expected value of position, 
thereby obtaining a classical equation of motion for that expected value:
\begin{equation}
m\ \frac{d^2\langle x\rangle}{dt^2} = - \frac{\partial V(\langle x
\rangle)}{\partial x}\ .
\label{onetwo} 
\end{equation}
This equation shows that the center of a narrow wave packet moves on an
orbit obeying Newton's laws.  More precisely, if we make a succession of
position and momentum measurements that are crude enough not to disturb
the approximation that allows (1.2) to replace (1.1), the expected values of
the results will be correlated by Newton's deterministic law.

This kind of elementary derivation is inadequate for the type of
classical behavior that we hope to discuss in quantum cosmology for the
following reasons:

\begin{itemize}
\item The behavior of expected or average values is not enough
to define classical behavior.  In quantum mechanics, the statement that
the moon moves on a classical orbit is properly the statement that, among
a set of alternative histories of its position as a function of time,
the probability is high for those histories exhibiting the correlations
in time implied by Newton's law of motion and near zero for all others.
To discuss classical behavior, therefore, we should be dealing with
the probabilities of individual time histories, not with expected or average
values.

\item The Ehrenfest theorem
 derivation deals with the results of ``measurements''
on an isolated system with a few degrees of freedom.  However, in
quantum cosmology we are interested in classical behavior in much more
general situations, over cosmological stretches of space and time, and
over a wide range of subsystems, {\it independent} of whether these
subsystems are receiving attention from observers.  Certainly we
imagine that our observations of the moon's orbit,  or a bit 
of the 
universe's expansion,
 have little to do with the classical behavior of those systems.
Further, we are interested not just in classical behavior as
exhibited in a few variables and at a few times of our choosing, but in
as  refined a description as possible, so that classical behavior
becomes a feature of the systems themselves and not a choice of
observers. 

\item The Ehrenfest theorem derivation relies on a close
connection between the equations of motion of the fundamental action and
the phenomenological deterministic laws that govern classical behavior.
But when we speak of the classical behavior of the
moon, or of the cosmological expansion, or even of water in a pipe, we
are dealing with systems with many degrees of freedom whose
phenomenological classical equations of motion may be only distantly
related to the underlying fundamental theory, say superstring theory.
We need a derivation which
derives the {\it form} of the equations as well as the probabilities that they
are satisfied.

\item The Ehrenfest theorem derivation posits the variables
--- the position $x$ --- in which classical behavior is exhibited.  But,
as mentioned above, classical behavior is most properly defined in terms
of the probabilities and properties of histories.  In a closed system we
should be able to {\it derive} the variables that enter into the
deterministic laws, especially because, for systems with many degrees of
freedom, these may be only distantly related to the co\"ordinates
entering the fundamental action.
\end{itemize}

Despite these shortcomings, the elementary Ehrenfest analysis
already exhibits two necessary requirements for classical behavior: Some
coarseness is needed
 in the description of the system as well as some restriction on its
initial condition. Not every initial wave function permits the
replacement of (\ref{oneone}) by (\ref{onetwo}) and therefore 
leads to classical behavior;
only for a certain class of wave functions will this be true.
Even given such a suitable  initial
condition, if we follow the system too closely, say by measuring
position exactly, thereby producing a completely delocalized state, we will
invalidate the approximation that allows (1.2) to replace (1.1)
 and classical behavior will not be
expected.  Some coarseness in the description of histories
 is therefore needed.  For realistic systems we therefore have the
important questions of {\it how restricted} is the class of initial
conditions which lead to classical behavior and {\it what} and {\it how
large} are the coarse grainings necessary to exhibit it.

Before pursuing these questions in the context of quantum cosmology I
would like to review a derivation of classical equations of
motion and the probabilities they are satisfied in a simple class of
model systems, but before doing {\it that} I must review, even more briefly,
the essential elements of the quantum mechanics of closed systems
\cite{Gri84,Omnsum,GH90a}.
 
\begin{figure}[t]
\begin{center}
{\epsfysize=2.00in \epsfbox{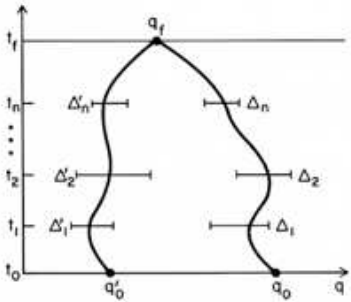}}
\caption{The sum-over-histories construction of the
decoherence functional.}
\end{center}
\end{figure}

\section{The Quantum Mechanics of Closed Systems}
\label{sec:II}

Most generally we aim at predicting the probabilities of alternative
time histories of a closed system such as the universe as a whole.
Alternatives at a moment of time are represented by an exhaustive set
of orthogonal projection operators $\{P^k_{\alpha_k} (t_k)\}$.  For
example, these might be projections on a set of alternative 
intervals for the center of mass position of a collection of particles,
or projections onto alternative ranges of their total momentum.  The
superscript denotes the set of alternatives
 \eg a certain set of  position ranges or a certain set of  momentum ranges, 
the discrete index $\alpha_k = 1,2,3 \cdots$ labels the
particular alternative, \eg a particular  range of position, 
and $t_k$ is the time.  A set of
alternative histories is defined by giving a series of such alternatives
at a sequence of times, say $t_1, \cdots, t_n$.  An individual history
is a sequence of alternatives $(\alpha_1, \cdots, \alpha_n)\equiv
\alpha$ and is represented by the corresponding chain of projections.
\begin{equation}
C_\alpha \equiv P^n_{\alpha_n} (t_n) \cdots P^1_{\alpha_1} (t_1)\ .
\label{twoone}
\end{equation}
Such a set is said to be ``coarse-grained'' because the $P$'s do not
restrict all possible variables and because they do not occur at all
possible times.

The decoherence functional 
\begin{equation}
D\left(\alpha^\prime, \alpha\right) = Tr\,\bigl[C_{\alpha^\prime} \rho
C^\dagger_\alpha\bigr]
\label{twotwo}
\end{equation}
measures the amount of quantum mechanical interference between pairs of
histories in a universe whose initial condition is  
represented by a density matrix $\rho$.
When, for a given set, the interference between all pairs of distinct
histories is sufficiently low, 
\begin{equation}
D\left(\alpha^\prime, \alpha\right) \approx  0\quad , \quad {\rm all}
\ \alpha^\prime \not=
\alpha 
\label{twothree}
\end{equation}
the set of alternative histories is said to {\it decohere}, and probabilities
can be consistently assigned to its individual members.  The probability
of an individual history $\alpha$ is just the corresponding 
diagonal element of $D$, {\it viz.}
\begin{equation}
p(\alpha) = D(\alpha, \alpha)\ . 
\label{twofour}
\end{equation}

Describe in terms of operators, check decoherence and evaluate
probabilities --- that is  how predictions are made for a closed system,
whether the alternatives are participants in a measurement situation or
not.

When the projections at each time  are onto the ranges $\{\Delta_\alpha\}$ of some generalized
co\"ordinates $q^i$ the decoherence functional can be written in a
convenient path integral from 
\begin{widetext}
\begin{equation}
D\left(\alpha^\prime, \alpha\right)  = \int_{\alpha^\prime}
\delta q^\prime \int_\alpha \delta q\, \delta \bigl(q^\prime_f
- q_f\bigr)
 e^{i(S[q^\prime(\tau)] - S[q(\tau)])/\hbar} \rho \left(q^\prime_0,
q_0\right) 
\label{twofive}
\end{equation}
\end{widetext}
where the integral is over the paths that pass through the intervals
defining the histories (Fig.~1). This form will be useful in what
follows.

\section{Classical Behavior in a Class of Model Quantum Systems}
\label{sec:III}

The class of models we shall discuss are defined by the following
features:

\begin{itemize}
\item We restrict attention to coarse grainings that follow a
fixed subset of the fundamental co\"ordinates $q^i$, say the center of mass
position of a massive body, and ignore the rest.  We denote the followed
variables by $x^a$ and the ignored ones by $Q^A$ so that  $q^i=(x^a, Q^A)$. We
thus posit, rather than derive, the variables exhibiting classical
behavior, but we shall derive, rather than posit,  the form of 
their phenomenological equations of motion.

\item We suppose the action is the sum of an action for the
$x$'s, an action for the $Q$'s, and an interaction between them that is
the integral of a local Lagrangian free from time derivatives.  That is,
\begin{equation}
S[q(\tau)] = S_{\rm free} [x(\tau)] + S_0 [Q(\tau)] + S_{\rm int}
[x(\tau), Q(\tau)] 
\label{threeone}
\end{equation}
suppressing indices where clarity is not diminished.

\item We suppose the initial density matrix factors into a
product of one depending on the $x$'s and another depending on the 
ignored $Q$'s which are
often called the ``bath'' or the ``environment''.
\end{itemize}

\begin{equation}
\rho\left(q^\prime_0, q_0\right) = \bar\rho \left(x^\prime_0,
x_0\right)\, \rho_B \left(Q^\prime_0, Q_0\right)\ . 
\label{threetwo}
\end{equation}
Under these conditions the integral  over the $Q$'s in (2.5) can be
carried out to give a decoherence functional just for coarse-grained
histories of the $x$'s of the form: 
\begin{widetext}
\begin{equation}
D\left(\alpha^\prime, \alpha\right)  =  \int_{\alpha^\prime} \delta x^\prime  
\int_{\alpha} \delta x\, \delta\bigl(x^\prime_f - x_f\bigr)
 \exp
\biggl\{i\Bigl(S_{\rm free} [x^\prime (\tau)]
 - S_{\rm free} [x(\tau)]
 + W
\left[x^\prime (\tau), x(\tau)\right]\Bigr)/\hbar\biggr\}\,
\bar\rho\left(x^\prime_0, x_0\right) 
\label{threethree}
\end{equation}
\end{widetext}
where $W [x^\prime(\tau), x(\tau)]$, called the 
Feynman-Vernon influence phase, summarizes the
results of integrations over the $Q$'s. 

\begin{figure}[t!]
\begin{center}
{\epsfysize=2.00in \epsfbox{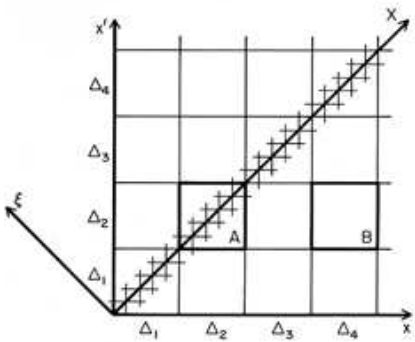}}
\caption{
The decoherence of histories coarse-grained by
intervals of a distinguished set of configuration space co\"ordinates.
The decoherence functional for such sets of histories is defined by the
double path integral of (\ref{threethree}) over paths $x^\prime(t)$ and
$x(t)$ that
are restricted by the coarse graining. These path integrals may be
thought of
as the limits of multiple integrals over the values of
$x^\prime$ and $x$ on a series of discrete time slices of the interval
$[0,T]$.  A typical slice at a time when the range of integration is
constrained by the coarse graining is illustrated. Of course, only
 one of the
distinguished co\"ordinates $x^a$ and its corresponding $x^{a\prime}$
can be shown and we have assumed for illustrative purposes that the
regions defining the coarse graining correspond to a set of intervals
$\Delta_\alpha, \alpha = 1, 2, 3, \cdots$ of this co\"ordinate.
On each slice where there is
a restriction from the coarse graining, the integration over $x^\prime$
and $x$ will be restricted to a single box.  For the ``off-diagonal''
elements of the decoherence functional corresponding to distinct
histories, that box will be off the diagonal (e.g. B) for {\it some}
slice.  For the diagonal elements, corresponding to the same histories,
the box will be on the diagonal (e.g. A) for all slices. \hfill\break
\indent If the imaginary part of the influence phase $W[x^\prime(\tau),
x(\tau)]$
grows as a functional of the magnitude of the 
difference $\xi(\tau) = x^\prime (\tau) -
x(\tau)$,
then the
integrand of the
decoherence functional will be negligible except when $x^\prime(\tau)$
is close to $x(\tau)$, a regime
illustrated by the shaded band about the diagonal in the
figure.  When the characteristic sizes of the intervals $\Delta_\alpha$
are large compared to the width of the band in which the integrand is
non-zero   the off-diagonal elements of the
decoherence functional will be negligible because integrals over those
slices where the histories are distinct is negligible (e.g.~over box B).
That is decoherence of the coarse-grained set of histories.  Further,
the evaluation of the diagonal elements of the decoherence
functional that give the probabilities of the individual histories in
decoherent set can be simplified.
If the integrations over $x^\prime$ and $x$ are
transformed to integrations over $\xi = x^\prime - x$ and $X=(x^\prime +
x)/2$ the restrictions on the range of the $\xi$-integration to one
diagonal box may be neglected with negligible error to the probability.}
\end{center}
\end{figure}

The influence phase $W$ generally possesses a positive imaginary part
\cite{Bru93}.
If that grows as $|x^\prime-x|$ increases, it will effect decoherence
because there will then be negligible contribution to the integral (3.3) for
$x^\prime \not= x$ or $\alpha^\prime \not= \alpha$. That, recall, is
the definition of decoherence (2.3).  Let us suppose this to be the case, as
is true in many realistic examples. Then we can make an important
approximation, which is a {\it decoherence} {\it expansion}.
Specifically, introduce co\"ordinates which measure the average and
difference between $x^\prime$ and $x$ (Fig.~2)
\begin{equation}
X = \half \left(x^\prime + x\right)\ , \quad \xi = x^\prime - x\ .
\label{threefour}
\end{equation}

The integral defining the diagonal elements of
$D$, which are the probabilities of the histories, receives a significant contribution
only for small $\xi(t)$.  We can thus
expand the exponent of the integrand of (\ref{threethree}) 
in powers of $\xi(t)$ and
legitimately retain only the lowest, say up to quadratic, terms.  The
result for the exponent is
\begin{widetext}
\begin{eqnarray}
S[x(\tau) & + & \xi(\tau)/2] - S[x(\tau) -\xi(\tau)/2] + W[x(\tau), \xi(\tau)]
\nonumber\\
& = & -\xi_0 P_0
 +  \int^T_0 dt\, \xi(t)\ \left[\frac{\delta S}{\delta X(t)} +
\left(\frac{\delta W}{\delta\xi(t)}\right)_{\xi(t)=0}\right]
 + \half \int^T_0 dt' \int^T_0 dt\ \xi(t^\prime)
\left(\frac{\delta^2 W}{\delta \xi(t')
\delta \xi(t)}\right)_{\xi(t)=0}
\ \xi(t) + \cdots\ . 
\label{threefive}
\end{eqnarray}
\end{widetext}
The essentially unrestricted integrals over the $\xi(t)$ can then be
carried out to give the following expression for the probabilities
\begin{widetext}
\begin{equation}
p(\alpha) = \int_\alpha \delta X\, ({\rm det}\ K_I/4\pi)^{-\half}
 \exp\Bigl[-\frac{1}{\hbar} \int^T_0 dt' \int^T_0 dt\ {\cal E} 
(t', X(\tau)]\,  K^{\rm inv}_I \left(t', t; X(\tau)\right]\ {\cal E}
(t, X(\tau)]\Bigr]\ \bar w \left(X_0, P_0\right)\ . 
\label{threesix}
\end{equation}
\end{widetext}
Here,
\begin{equation}
{\cal E}(t, X(\tau)] \equiv  \frac{\delta S}{\delta X(t)} +
\left\langle F(t, X(\tau)]\right\rangle 
\label{threeseven}
\end{equation}
where $\langle F(t, X (\tau)]\rangle$ has been written for $(\delta
W/\delta\xi(t))_{\xi=0}$ because it can be shown to be the expected value of the
force arising from the ignored variables in the state of the bath.  $K^{\rm
inv}(t', t, X(\tau)]$ is the inverse of $(2\hbar/i)(\delta^2
W/\delta\xi(t')\delta\xi(t))$ which turns out to be real and
positive.  Finally $\bar w(X,P)$ is the Wigner distribution for the density
matrix $\bar\rho$:
\begin{equation}
w(X, P) = \frac{1}{2\pi} \int d\xi\, e^{i P\xi /\hbar}\rho (X+\xi/2, X-
\xi/2)\ .
\label{threeeight}
\end{equation}
This expression shows that, when $K^{\rm inv}_I$ is sufficiently large,
the probabilities for histories of  $X(t)$ are peaked about those which
satisfy the equation of motion
\begin{equation}
{\cal E} (t, X(\tau)] = \frac{\delta S}{\delta X(t)} + \langle F(t,
X(\tau)]\rangle = 0\ . 
\label{threenine}
\end{equation}
and the initial conditions of these histories are distributed 
according to the Wigner
distribution.  The Wigner distribution is not generally positive, but,
up to the accuracy of the approximations, this integral of it must be
\cite{Hal92}.  

Thus we derive the form of the phenomenological equations of motion for
this class of models.  It is the equation of motion of the fundamental action
$S[X(t)]$ corrected by phenomenological forces arising from the
interaction with the bath.  These depend not only on the form of the
interaction Hamiltonian but also on the initial state of the bath,
$\rho_B$.  These forces are generally
non-local in time, depending at a given instant on the whole trajectory
$X(\tau)$. It can be shown that quantum mechanical causality implies that
they depend only on part of path $X(\tau)$ to the past of $t$.  Thus
quantum mechanical causality implies classical causality.

It is important to stress that the expansion of the decoherence
functional has enabled us to consider the
equations of motion for fully non-linear systems, not just the linear
oscillator models that have been widely studied.

The equation of motion (\ref{threenine}) is not predicted to be satisfied 
{\it exactly}.  The
probabilities are {\it peaked} about ${\cal E}=0 $ but distributed
about that value with a width that depends on the size of $K^{\rm inv}$.
That is quantum noise whose spectrum and properties can be derived from 
(\ref{threethree}).  The fact that both the spectrum of fluctuations
and the phenomenological forces can be derived from the same influence
phase is the origin of the fluctuation dissipation theorem for linear
systems. 

Simple examples of this analysis are the linear oscillator models that have
been studied
using path integrals by
Feynman and Vernon \cite{FV63}, Caldeira and Leggett \cite{CL83}, 
Unruh and Zurek \cite{UZ89}, 
and many
others.  For these, the $x$'s describe a distinguished 
harmonic oscillator linearly
coupled to a bath of many others.  If the initial state of the bath is a
thermal density matrix, then the decoherence expansion is 
exact. In the especially simple case of a cut-off continuum of bath
oscillators and high bath temperature, there  are
the following results:  The imaginary part of the influence phase is
given by
\begin{equation}
ImW[x'(\tau),x(\tau)]= \frac{2M\gamma kT_B}{\hbar} \int^T_0 dt 
\left(x^\prime(t) -
x(t)\right)^2 
\label{threeten}
\end{equation}
where $M$ is the mass of the $x$-oscillator, $\gamma$ is a measure of the
strength of its coupling to the bath, and $T_B$ is the temperature of the
bath.  The exponent of the expression (\ref{threeseven}) 
giving the probabilities for
histories is
\begin{equation}
-\frac{M}{8\gamma kT_B} \int^T_0 dt\, \left[\ddot X + \omega^2  X +
2\gamma \dot X\right]^2 
\label{threeeleven}
\end{equation}
where $\omega$ is the frequency of the $x$-oscillator renormalized by its
interaction with the bath. The phenomenological force is 
friction, and the occurrence of $\gamma$, both in that force and the
constant in front of (\ref{threeeleven}), whose size governs the 
deviation from classical
predictability, is a simple example of the fluctuation-dissipation theorem.

In this simple case, an analysis of the the
requirements for classical behavior is straightforward.  
To achieve decoherence we need
high values of $\gamma kT_B$.  That is, strong coupling is needed if
interference phases are to be dissipated efficiently into the bath.
However, the larger the value of $\gamma kT_B$ the smaller the
coefficient of front of (\ref{threeeleven}), decreasing the size of the 
exponential
and {\it increasing} deviations from classical predictability.  This is
reasonable: the stronger the coupling to the bath the more noise
is produced by the interactions that are carrying away the phases.  To
counteract that, and achieve a sharp peaking about the classical
equation of motion, $M$ must be large so that $M/\gamma kT_B$ is large.
That is, high inertia is needed to resist the noise that arises from the
interactions with the bath.

Thus, much more coarse graining is needed to ensure classical
predictability than naive arguments based on the uncertainty principle
would suggest.  Coarse graining is needed to effect decoherence, and
coarse graining beyond that to achieve the inertia necessary to resist
the noise that the mechanisms of decoherence produce.

\section{Quasiclassical Realms in Quantum Cosmology}
\label{sec:IV}

As observers of the universe, we deal every day with coarse-grained
histories that exhibit classical correlations.  Indeed, only by
extending our direct perceptions with expensive and delicate instruments
can we exhibit {\it non}-classical behavior. The coarse grainings that
we use individually and collectively are, of course, characterized by a
large amount of ignorance, for our observations determine only a very few
of the variables that describe the universe and those only very
imprecisely.  Yet, we have the impression that the universe exhibits a
much finer-grained set of histories, {\it independent of our choice},
defining an always decohering ``quasiclassical realm'', to which our
senses are adapted but deal with only a small part of. If we are
preparing for a journey to a yet unseen part of the universe, we do not
believe that we need to equip our spacesuits with detectors, say
sensitive to coherent superpositions of position or other unfamiliar
quantum operators.  We expect that histories of familiar quasiclassical
operators will decohere and exhibit patterns of classical correlation
there as well as here. 

Roughly speaking, a quasiclassical realm is a set of decohering
histories, that is maximally refined
with respect to decoherence, and whose individual histories exhibit as much
as possible patterns of deterministic correlation.  At present we lack
satisfactory measures of maximality and classicality with which to make
the existence of one or more quasiclassical realms into  quantitative
questions in quantum cosmology \cite{GH90a,PZ93}. 
We therefore do not know whether the
universe exhibits a {\it unique} class of roughly equivalent sets of histories
with high levels of classicality constituting the quasiclassical realm 
of familiar experience, or whether there might be other essentially
inequivalent quasiclassical realms \cite{GH94}.  
However, even in the absence of such
measures and such analyses, we can make an argument for the 
form of at least some of the operators we expect to occur over and over
again in histories defining one kind of quasiclassical realm --- operators we
might call ``quasiclassical''. In the earliest instants of the history of
the universe, the coarse grainings
defining spacetime geometry on scales above the Planck scale must 
emerge as quasiclassical.  Otherwise,
our theory of the initial condition is simply inconsistent with
observation in a manifest way. Then, when there is classical spacetime
geometry we can consider the conservation of energy, and momentum, and
of other quantities which are conserved by virtue of the equations of
quantum fields.  Integrals of densities of conserved or nearly conserved
quantities over suitable volumes are natural candidates for
quasiclassical operators.  Their approximate conservation allow them to
resist deviations from predictability caused by ``noise'' arising from
their interactions with the rest of the universe that accomplish
decoherence.  Such ``hydrodynamic'' variables {\it are} among the
principal variables of classical theories.

This argument is not unrelated to a standard one in classical 
statistical mechanics that seeks to identify the variables in which a hydrodynamic description
of non-equilibrium systems may be expected. 
All isolated systems approach equilibrium --- that is statistics. With
certain coarse grainings this approach to equilibrium may be 
approximately described by hydrodynamic equations, such as the Navier-Stokes
equation, incorporating phenomenological descriptions of dissipation,
viscosity, heat conduction, diffusion, etc. The variables that
characterize such hydrodynamic descriptions are the local quantities
which very most {\it slowly} in time --- that is, averages of
densities of approximately conserved quantities over suitable volumes.
The volumes must be large enough that statistical fluctuations
in the values of the averages are small, but small enough that equilibrium
is established within each volume in a time short compared to the dynamical
times on which the variables vary. The constitutive relations
defining coefficients of viscosity, diffusion, etc. are then defined
and independent of the initial condition, permitting the closure
of the set of hydrodynamic equations. Local equilibrium being
established, the further equilibration of the volumes among themselves is
described by the hydrodynamic equations. In the context of quantum
cosmology, coarse grainings by averages of densities of approximately conserved quantities
not  only permit local equilibrium and resist gross statistical fluctuations leading to high probabilities for deterministic histories as in this argument, 
they also, as described above,  resist the fluctuations arising from the mechanicsms
of decoherence necessary for predicting probabilities of any kind
in quantum mechanics.

In this way we can sketch how a quasiclassical realm 
consisting of histories of ranges of values of quasiclassical operators, extended over cosmological dimensions both in
space and in time, but highly refined with respect to those scales, is
a feature of our universe and thus must be a prediction of
its quantum initial condition.  It may seem strange to attribute the
classical behavior of everyday objects to the initial condition of the
universe some 12 billion years ago, but, in this connection, two things
should be noted: First, we are not just speaking of the classical behavior
of a few objects described in a very coarse graining of our choosing,
but of  a much more refined feature of the universe extending over
cosmological dimensions and indeed including the classical behavior of
the cosmological geometry itself all the way back to the briefest of
moments after the big bang. Second, at the most fundamental level 
the {\it only} ingredients entering into quantum mechanics are the theory of the initial condition and the theory of dynamics,
so that {\it any} feature of the universe must be traceable to these two starting
points and the accidents of our particular history.  Put differently (neglecting quantum gravity)  
the possible classical
behavior of a set of histories represented by strings of projection
operators as in (\ref{twoone}) does not depend on the operators 
alone except in
trivial cases.  Rather, like decoherence itself, classicality depends on
the relation of those operators to the initial state $|\Psi\rangle$
through which we calculate the decoherence and probabilities of
sets of histories by which classical behavior is defined.

Yet it is reasonable to ask --- how sensitive is the existence of a
quasiclassical realm to the particular form of the initial condition?
In seeking to answer this question it is important to recognize that
there are two things it might mean.  First, we might ask whether {\it
given} an initial state $|\Psi\rangle$, there is always a set of
histories which decoheres and exhibits deterministic correlations.
There is, trivially. Consider the set of histories which just consists
of projections down on ranges $\{\Delta E_\alpha\}$ of the {\it total}
energy (or any other conserved quantity) at a sequence of times
\begin{equation}
C_\alpha = P^H_{\alpha_n} (t_n) \cdots P^H_{\alpha_1} (t_1) 
\ .\label{fourone}
\end{equation}
Since the energy is conserved these operators are independent
of time, commute, and $C_\alpha$ is merely the projection onto the
intersection of the intervals 
$\Delta E_{\alpha_`}, \cdots, \Delta
E_{\alpha_n}$.
The set of histories represented by (\ref{fourone}) thus 
{\it exactly} decoheres
\begin{equation}
D\left(\alpha^\prime, \alpha\right) = Tr\bigl[C_{\alpha^\prime}
|\Psi\rangle\langle\Psi|C^\dagger_\alpha\bigr] =
\langle\Psi|C^\dagger_\alpha C_{\alpha^\prime} |\Psi \rangle \propto
\delta_{\alpha\alpha^\prime}\ ,
\label{fourtwo}
\end{equation}
and exhibits deterministic correlations --- the total energy today is the same
as it was yesterday. Of course, such a set is far from maximal, but
imagine subdividing the total volume again and again and considering the
set of histories which results from following the values of the energy
in each subvolume over the sequence of times.  If the process of
subdividing is followed until we begin to lose decoherence we might hope to
retain some level of determinism while moving towards maximality.  Thus,
it seems likely that, for most initial $|\Psi\rangle$, we may find {\it some}
sets of histories which constitute a quasiclassical realm.

However, we might ask about the sensitivity of a
quasiclassical realm to initial condition in a different way. We might fix
the chains of projections
that describe {\it our} highly refined quasiclassical realm and ask for
how many {\it other} initial states does this set of histories decohere
and exhibit the same classical correlations.  This amounts to asking, for
a given set of alternative histories $\{C_\alpha\}$, how many
initial states $|\Psi\rangle$ will have the same decoherence functional? Expand
$|\Psi\rangle$ in some generic basis in Hilbert space, $|i\rangle$:
\begin{equation}
|\Psi\rangle = \sum\nolimits_i c_i | i\rangle\ .
\label{fourthree}
\end{equation}
The condition that $|\Psi\rangle$ result in a given decoherence
functional $D(\alpha^\prime, \alpha)$ is
\begin{equation}
\sum\nolimits_{ij} c^*_i c_j\ \bigl\langle i|C^\dagger_{\alpha^\prime}
C_\alpha | j\bigr\rangle = D\left(\alpha^\prime, \alpha\right)\ .
\label{fourfour}
\end{equation}
Unless the $C_\alpha$ are such that decoherence and correlations are
trivially implied by the operators (as is the above example of chains of
projections onto a total conserved energy), the matrix elements $\langle
i|C^\dagger_{\alpha^\prime}\, C_\alpha | j\rangle$ will not vanish
indentically.  Equation (\ref{fourfour}) is therefore (number of histories
$\alpha)^2$ equations for (dimension of Hilbert space) coefficients.
When that dimension is made finite, say by limiting the total volume and
energy, we expect a solution only when
\begin{equation}
\left({\rm number\ of\ histories}\atop{\rm in\ the\ quasiclassical
\ realm}\right)^2 \ltwid \left({\rm dim}
\ {\cal H}\right)\ .
\label{fourfive}
\end{equation}
As the set of histories becomes increasingly refined, so that there are
more and more alternative cases, the two sides may come closer to
equality.  The number of states $|\Psi\rangle$ which reproduce the
{\it particular} maximal quasiclassical realm of our universe may thus be
large but still small compared to the total number of states in Hilbert
space.

\section{The Main Points Again}
\label{sec:V}

\begin{itemize}
\item Classical behavior of quantum systems is defined
through the probabilities of deterministic correlations of individual
time histories of a closed system.
\item Classical predictability requires coarse graining to
accomplish decoherence, and coarse graining beyond that to achieve the
necessary inertia to resist the noise which mechanisms of decoherence
produce.
\item The maximally refined quasiclassical realm of familiar
experience is an emergent feature, not of quantum evolution alone, but of
that evolution, coupled to a specific theory of the universe's initial
condition.  Whether the whole closed system exhibits a quasiclassical
realm like ours, and indeed whether it exhibits more than one
essentially inequivalent realm, are calculable questions in quantum
cosmology if suitable measures of maximality and classicality can be
supplied.
\item A generic initial state will exhibit some sort of
quasiclassical realm, but the maximally refined quasiclassical realm of
familiar experience will be an emergent feature of only a small fraction
of the total possible initial states of the universe.
\end{itemize}

\acknowledgments

Most of this paper reports joint work with M.~Gell-Mann. The author's
research was supported in part by NSF grant PHY90-08502.

\end{document}